\documentclass[%
reprint,
nofootinbib,
nobibnotes,
amsmath,amssymb,
aps,
]{revtex4-2}

\usepackage{graphicx}
\usepackage{dcolumn}
\usepackage{bm}

\usepackage{amsmath,amssymb,amsbsy,amstext,amsthm,simplewick,amsfonts}
\usepackage{mathrsfs}
\usepackage{graphicx}
\usepackage{wrapfig}
\usepackage{upgreek}
\usepackage{bm} 
\usepackage{framed}
\usepackage{bbm}
\usepackage{textcomp}
\usepackage{adjustbox}
\usepackage{makecell}
\usepackage{tcolorbox}
\usepackage{empheq}
\usepackage[normalem]{ulem}
\usepackage{enumitem}
\usepackage{braket}
\usepackage{array}
\usepackage{dsfont}
\usepackage{physics}
\usepackage{ulem}
\usepackage{xcolor}
\usepackage{tocloft}
\usepackage{ulem}

\usepackage{orcidlink}

\usepackage{hyperref}
\hypersetup{colorlinks=true,linkcolor=bluecyan,citecolor=red3,urlcolor=green3,pdfencoding=auto,linktocpage
}
\makeatletter
\let\pre@bibdata\@empty
\let\@bibdataout@init\relax
\makeatother

\definecolor{bluecyan}{RGB}{0, 100, 200}
\definecolor{red3}{RGB}{214,39,40}
\definecolor{green3}{RGB}{44,160,44}

\def \dd{\mathrm{d}}

\newmuskip\pFqmuskip
\newcommand{\pFqcomma}{\mathchar"613B\mskip\pFqmuskip}
\newcommand{\pFq}[6][8]{
  \begingroup
  \pFqmuskip=#1mu\relax
  \mathcode`\,=\string"8000
  \begingroup\lccode`\~=`\,
  \lowercase{\endgroup\let~}\pFqcomma
  {}_{#2}F_{#3}\left[\genfrac..{0pt}{}{#4}{#5};#6\right]
  \endgroup}

\begin{document}
\title{Cosmological Correlators with Triple Massive Exchanges}
\author{Shuntaro Aoki\textsuperscript{1}}
\email{shuntaro.aoki@riken.jp}
\author{Lucas Pinol\textsuperscript{2}}
\email{lucas.pinol@phys.ens.fr}
\author{Zhehan Qin\textsuperscript{3}}
\email{zq242@cam.ac.uk}
\author{Fumiya Sano\textsuperscript{4}}
\email{sanof.cosmo@gmail.com}
\author{Masahide Yamaguchi\textsuperscript{5,6,7}}
\email{gucci@ibs.re.kr}
\author{Yuhang Zhu\textsuperscript{5}}
\email{yhzhu@ibs.re.kr}

\affiliation{\vspace{6pt}\textsuperscript{1}RIKEN Center for Interdisciplinary Theoretical and Mathematical Sciences (iTHEMS), Wako, Saitama 351-0198, Japan}
\affiliation{\vspace{3pt}\textsuperscript{2}Laboratoire de Physique de l'\'Ecole Normale Sup\'erieure,
ENS, CNRS, Universit\'e PSL,
Sorbonne Universit\'e, Universit\'e Paris Cit\'e,
F-75005 Paris, France}
\affiliation{\vspace{3pt}\textsuperscript{3}Department of Applied Mathematics and Theoretical Physics, University of Cambridge, Wilberforce Road, Cambridge, CB3 0WA, UK}
\affiliation{\vspace{3pt}\textsuperscript{4}Graduate School of Arts and Sciences, the University of Tokyo, 3-8-1 Komaba, Meguro-ku, Tokyo 153-8902, Japan}
\affiliation{\vspace{3pt}\textsuperscript{5}Cosmology, Gravity and Astroparticle Physics Group, Center for Theoretical Physics of the Universe, Institute for Basic Science, Daejeon 34126, Korea}
\affiliation{\vspace{3pt}\textsuperscript{6}Department of Physics, Institute of Science Tokyo, 2-12-1 Ookayama, Meguro-ku, Tokyo 152-8551, Japan}
\affiliation{\vspace{3pt}\textsuperscript{7}Department of Physics and IPAP, Yonsei University, 50 Yonsei-ro, Seodaemun-gu, Seoul 03722, Korea}

\date{\today}
\begin{abstract}
Among all tree-level contributions to the primordial bispectrum, the triple exchange of massive scalars is the most challenging. In this paper we present an analytical formula for this bispectrum in the form of a double series, which allows the three exchanged fields to have different masses and converges throughout the whole physical region. We obtain it by expanding the mixed propagator with the softest momentum, so that the triple exchange becomes a sum of integrals over two mixed propagators. Each of these integrals satisfies a differential equation and is fixed by its value at equal momentum. We further show that the squeezed limit follows directly from this formula and takes a simple and compact form.
\end{abstract}
\maketitle

\section{Introduction}
\label{sec:introduction}

Primordial correlators can carry traces of heavy particles present during inflation. In particular, the exchange of such a particle between inflaton fluctuations gives the correlator a nonanalytic kinematic dependence, from which its mass can be read off~\cite{Chen:2009zp,Baumann:2011nk,Noumi:2012vr,Arkani-Hamed:2015bza}. Extracting it quantitatively requires accurate theoretical predictions for this kinematic dependence.

In most cases, the relevant predictions are those for tree-level correlators, which typically give the leading contribution to primordial non-Gaussianity. They also link the parameters of a model to the observed signal more directly, free of the ultraviolet renormalisation subtleties of loop corrections. Analytical methods for cosmological correlators have developed rapidly over the past few years and now cover a large class of tree-level processes with massive exchanges \cite{Arkani-Hamed:2018kmz, Baumann:2019oyu,Sleight:2019mgd,Sleight:2019hfp,Sleight:2020obc,Pimentel:2022fsc,Jazayeri:2022kjy,Qin:2022fbv,Werth:2024mjg,Chen:2024glu,Liu:2024str,Stefanyszyn:2024msm,Qin:2025xct,Meltzer:2021zin,Sleight:2021plv,Chen:2023iix,Liu:2024xyi,Xianyu:2025lbk,Jazayeri:2025vlv,Belrhali:2026ygh,Belrhali:2026jqe,Gasparotto:2024bku,
Cheung:2025dmc,Belrhali:2026ktb,Baumann:2026atn,Grafe:2026dqn,Aoki:2023wdc}. Single \cite{Arkani-Hamed:2018kmz,Qin:2023ejc} and double massive exchanges \cite{Aoki:2024uyi,Xianyu:2023ytd,Belrhali:2026act,Grafe:2026avi} are now known analytically beyond the squeezed limit, which makes their full kinematic dependence accessible. Tree-level correlators can also be evaluated numerically and systematically, with the cosmological flow approach and its implementation in \texttt{CosmoFlow} \cite{Werth:2023pfl,Pinol:2023oux,Werth:2024aui}. These results have been used to construct templates for cosmological collider searches \cite{Cabass:2024wob,Sohn:2024xzd,Suman:2025tpv,Kumar:2026dih,Philcox:2025wts,Green:2026yev,Philcox:2026tjj,Kumar:2026ogn,Suman:2025vuf}.

Quasi-single field inflation \cite{Chen:2009zp,Chen:2009we} provides an early and influential example of a massive field generating sizeable tree-level non-Gaussianity. In this model, the leading contribution to the bispectrum comes from the triple exchange diagram, in which a cubic self-interaction of the massive scalar is connected to three external inflaton fluctuations through quadratic mixing. Although three mixing insertions are required, this contribution can be large, since the self-coupling of the massive field need not be suppressed by the slow-roll parameters. Also with single and double exchanges now under analytical control, the full shape of the triple exchange is the natural and important next target.

Very recently, the two-field system of the inflaton and the massive field was solved analytically for an arbitrary mixing strength \cite{Huenupi:2026abj}, and the corresponding bispectrum has since been studied with a variety of methods \cite{Huenupi:2026aqc,Pinol:2026hey,Pinol:2026xnl,Belrhali:2026uxn,Wang:2026lff}. In the weak mixing limit, the part of this bispectrum generated by the cubic self-interaction reduces to the triple exchange of a single massive field. The coefficient of its leading squeezed limit was obtained in closed form \cite{Pinol:2026hey}. The exact solution has also been extended to unequal sound speeds \cite{Huenupi:2026jcv} and to spin-2 fields with the chemical potential \cite{Gorji:2026rvx}.

With more than one massive species, the squeezed bispectrum can display distinctive features \cite{Emami:2013lma,Pinol:2021aun,Aoki:2020zbj}. In this paper, we work in the weak mixing regime and derive the analytical expression for the triple massive exchange bispectrum generated by a cubic interaction among three massive scalars with independently specified masses. By expanding the mixed propagator that carries the smallest external momentum, we reduce the calculation to integrals over two mixed propagators. Each of these integrals satisfies an inhomogeneous differential equation and is fixed by regularity together with its value at equal momentum which we determine analytically. The resulting series involves generalised hypergeometric functions of a single variable and converges throughout the physical region, including equilateral, squeezed and folded configurations.

The paper is organised as follows. In Sec.~\ref{sec:setup}, we introduce the model and reduce the bispectrum to a single time integral over three mixed propagators. Sec.~\ref{sec:solution} derives our analytical result, and Sec.~\ref{sec:kinematics} discusses its squeezed limit and the equal mass case. We summarise our results in Sec.~\ref{sec:summary} and extend them to arbitrary positive masses in App.~\ref{app:complementary}.

\vspace{2pt}
\noindent\textbf{Conventions.} Throughout this paper, the background is four-dimensional inflationary spacetime, $ds^2=a^2(\tau)(-d\tau^2+d\boldsymbol x^2)$, with $a(\tau)=-1/(H\tau)$ and $-\infty<\tau<0$. We denote the comoving momentum by $\boldsymbol k_i$, with magnitude $k_i\equiv|\boldsymbol k_i|$, and abbreviate their sums as, e.g., ${k_{123}\equiv k_1+k_2+k_3}$. Primes denote derivatives with respect to conformal time, except on correlation functions, where they indicate that the overall momentum-conserving delta function $(2\pi)^3\delta^{(3)}(\sum_i\boldsymbol k_i)$ has been removed.

\section{Mixed propagators and\\ ~~~~~~seed integral}
\label{sec:setup}

\begin{figure}[htbp]
\centering
\includegraphics[width=\columnwidth]{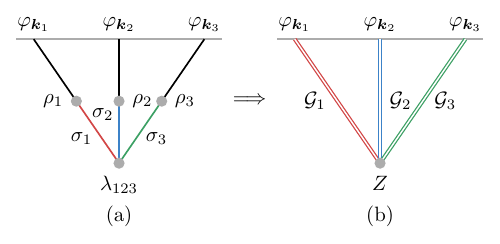}
\caption{The triple exchange diagram and its reduction. (a) The inflaton fluctuation (black lines) mixes with the massive fields $\sigma_\alpha$ (coloured lines) at the vertices $\rho_\alpha$, and the three massive fields meet at the cubic vertex $\lambda_{123}$. The horizontal line represents the late-time boundary at $\tau=0$, and each vertex carries an SK index $\pm$ to be summed over. (b) Once we sum over the SK indices, each mixing vertex is absorbed into a mixed propagator $\mathcal G_\alpha$ (double lines). The diagram then reduces to a single time integral over the cubic vertex.}
\label{fig:triple_exchange}
\end{figure}

In general, the massive fields exchanged in the triple exchange diagram need not be the same. We therefore consider $N$ massive scalars $\sigma_\alpha$ with masses $m_\alpha$, together with the massless inflaton fluctuation $\varphi$, all with canonical kinetic terms and unit sound speed. With repeated indices summed over, their interactions read
\begin{align}
S_{\rm int}&=\int\dd\tau\,\dd^3x\,\Big[a^3\rho_\alpha\,\varphi'\sigma_\alpha-\frac{a^4}{3!}\lambda_{\alpha\beta\gamma}\,\sigma_\alpha\sigma_\beta\sigma_\gamma\Big]\,, \label{action}
\end{align}
$\rho_\alpha$ and $\lambda_{\alpha\beta\gamma}$ are real constant couplings, the latter being totally symmetric. For $N=1$, this reduces to the cubic self-interaction of quasi-single field inflation \cite{Chen:2009zp}. In perturbation theory, the quadratic mixings are treated as interactions, and the fields are described by their free mode functions in the Bunch-Davies vacuum,
\begin{align}
\varphi(k,\tau)&=\frac{H}{\sqrt{2k^3}}(1+ik\tau)e^{-ik\tau}\,,\nonumber\\
\sigma_\alpha(k,\tau)&=\frac{\sqrt\pi}{2}\,e^{-\pi\mu_\alpha/2-i\pi/4}H(-\tau)^{3/2}H^{(1)}_{i\mu_\alpha}(-k\tau)\,, \label{mode_functions}
\end{align}
with $\mu_\alpha\equiv\sqrt{m_\alpha^2/H^2-9/4}$. In the main text we focus on the case that all scalars lie in the principal series, namely $\mu_\alpha>0$, while our result can be easily extended to general massive case, as presented in the App.~\ref{app:complementary}. The corresponding bulk-to-bulk propagators of the massive fields in the Schwinger-Keldysh (SK) formalism are \cite{Chen:2017ryl}
\begin{align}
D^{(\alpha)}_{-+}(k;\tau_1,\tau_2)&=\sigma_\alpha(k,\tau_1)\,\sigma^*_\alpha(k,\tau_2)\,,\nonumber\\
D^{(\alpha)}_{+-}(k;\tau_1,\tau_2)&=\sigma^*_\alpha(k,\tau_1)\,\sigma_\alpha(k,\tau_2)\,,\nonumber\\
D^{(\alpha)}_{++}(k;\tau_1,\tau_2)&=\theta(\tau_1-\tau_2)D^{(\alpha)}_{-+}+\theta(\tau_2-\tau_1)D^{(\alpha)}_{+-}\,,\nonumber\\
D^{(\alpha)}_{--}(k;\tau_1,\tau_2)&=\theta(\tau_1-\tau_2)D^{(\alpha)}_{+-}+\theta(\tau_2-\tau_1)D^{(\alpha)}_{-+}\,. \label{bulk_propagators}
\end{align}
At leading order in the couplings, the bispectrum receives contributions from the triple exchange diagrams of the form shown in Fig.~\ref{fig:triple_exchange},
\begin{align}
\langle\varphi_{\boldsymbol k_1}\varphi_{\boldsymbol k_2}\varphi_{\boldsymbol k_3}\rangle'&=\sum_{\alpha,\beta,\gamma}\mathcal B_{\alpha\beta\gamma}(k_1,k_2,k_3)\,, \label{full_bispectrum}
\end{align}
with $\mathcal B_{\alpha\beta\gamma}$ the diagram in which $\sigma_\alpha$, $\sigma_\beta$ and $\sigma_\gamma$ are attached to $\varphi_{\boldsymbol k_1}$, $\varphi_{\boldsymbol k_2}$ and $\varphi_{\boldsymbol k_3}$, respectively. The sum over species labels thus automatically accounts for the permutations of the external momentum. Since all these diagrams take the same form, we only need to compute the diagram $\mathcal B(k_1,k_2,k_3)\equiv\mathcal{B}_{123}$ shown in Fig.~\ref{fig:triple_exchange}, with its internal lines labelled by $1,2,3$. Any other diagram $\mathcal B_{\alpha\beta\gamma}$ then follows by replacing $\mu_1,\mu_2,\mu_3$ with $\mu_\alpha,\mu_\beta,\mu_\gamma$, and $\lambda_{123}\rho_1\rho_2\rho_3$ with $\lambda_{\alpha\beta\gamma}\rho_\alpha\rho_\beta\rho_\gamma$.

Labelling the SK indices of the three mixing vertices by $\mathsf a_j=\pm$, with $j=1,2,3$, and that of the cubic vertex by $\mathsf b=\pm$, the diagram takes the form
\begin{align}
\mathcal B(k_1,k_2,k_3)&=\frac{\lambda_{123}\rho_1\rho_2\rho_3}{8H^7k_1k_2k_3}\sum_{\mathsf a_1,\mathsf a_2,\mathsf a_3,\mathsf b=\pm}\mathsf b\int_{-\infty}^0\frac{\dd\tau_4}{\tau_4^4}\nonumber\\
&\times\prod_{j=1}^3\mathsf a_j\int_{-\infty}^0\frac{\dd\tau_j}{\tau_j^2}\,e^{i\mathsf a_jk_j\tau_j}D^{(j)}_{\mathsf a_j\mathsf b}(k_j;\tau_j,\tau_4)\,. \label{sk_bispectrum}
\end{align}
Note that, at fixed $\tau_4$ and $\mathsf b$, the three mixing vertices in \eqref{sk_bispectrum} are independent of each other. We can therefore perform the time integral and the SK sum at each mixing vertex separately, which leads to the mixed propagator \cite{Chen:2017ryl}
\begin{align}
\mathcal G^{(\alpha)}_{\mathsf b}(k;\tau_4)&\equiv\sum_{\mathsf a=\pm}\mathsf a\int_{-\infty}^0\frac{\dd\tau}{\tau^2}\,e^{i\mathsf ak\tau}D^{(\alpha)}_{\mathsf{ab}}(k;\tau,\tau_4)\,. \label{mixed_def}
\end{align}
With \eqref{mixed_def}, the bispectrum \eqref{sk_bispectrum} reduces to a single time integral over the cubic vertex. In practice, it is convenient to Wick rotate the time contours following the Bunch-Davies prescription. Since the $\mathsf b=-$ contribution is the complex conjugate of the $\mathsf b=+$ one, it suffices to rotate $\tau_4$ to the positive imaginary axis, along which the mixed propagator depends on $k$ and $\tau_4$ only through ${y\equiv-ik\tau_4>0}$,
\begin{align}
\mathcal G^{(\alpha)}_+(k;\tau_4)&=-\frac{iH^2}{k^2}\,y^{3/2}\,\mathcal G_\alpha(y)\,. \label{mixed_wick}
\end{align}
Introducing $Z\equiv-ik_{123}\tau_4/2$ and the momentum ratios $r_\alpha\equiv2k_\alpha/k_{123}$, we then obtain
\begin{align}
\mathcal B(k_1,k_2,k_3)&=\frac{\lambda_{123}\rho_1\rho_2\rho_3}{\sqrt2\,H(k_1k_2k_3k_{123})^{3/2}}\,\mathrm{Re}\,\mathcal I\,, \label{bispectrum_seed}
\end{align}
with the seed integral
\begin{align}
\mathcal I&\equiv\int_0^\infty\dd Z\,Z^{1/2}\prod_{\alpha=1}^3\mathcal G_\alpha(r_\alpha Z)\,. \label{seed_integral}
\end{align}
The momentum ratios satisfy $r_1+r_2+r_3=2$ and lie in $0<r_\alpha<1$ for physical triangles, and $r_\alpha=1$ at folded configurations. The time integral in \eqref{mixed_def} can be carried out explicitly, then the mixed propagator reads \cite{Chen:2017ryl}
\begin{align}
\mathcal G_\alpha(y)&=\beta_\alpha\bigg[\frac{\pi\,I_{i\mu_\alpha}(y)}{1+ie^{-\pi\mu_\alpha}}-K_{i\mu_\alpha}(y)\bigg]\nonumber\\
&\quad-\frac{\sqrt y\,e^{-y}}{1/4+\mu_\alpha^2}\,\pFq{2}{2}{1,1}{\frac32+i\mu_\alpha,\frac32-i\mu_\alpha}{2y}\,, \label{mixed_closed}
\end{align}
where $\beta_\alpha\equiv\sqrt{\pi/2}\,(1+ie^{-\pi\mu_\alpha})/\cosh(\pi\mu_\alpha)$, $I_\nu$ and $K_\nu$ are the modified Bessel functions, and ${}_pF_q$ is the generalised hypergeometric function. Acting on the mixed propagator with $\mathcal D_\alpha\equiv y^2\partial_y^2+y\partial_y-y^2+\mu_\alpha^2$ turns the bulk-to-bulk propagator into a delta function, which gives
\begin{align}
\mathcal D_\alpha\mathcal G_\alpha(y)&=-y^{1/2}e^{-y}\,. \label{mixed_eom}
\end{align}
As expected, the square bracket in \eqref{mixed_closed} solves the homogeneous equation, while the ${}_2F_2$ term provides a particular solution.

\section{Analytical solution}
\label{sec:solution}

Let us now see how the integral \eqref{seed_integral} can be computed. The product of three mixed propagators cannot be integrated directly, but the problem becomes much simpler once we expand the mixed propagator with the smallest momentum, since the bispectrum can then be written in terms of integrals of only the other two. It turns out that, thanks to \eqref{mixed_eom}, these integrals satisfy an inhomogeneous differential equation, and its solution is fixed by a single boundary value at $k_2=k_3$. This boundary value can in turn be computed using the Laplace representation of the mixed propagators. We now go through these steps one by one.

\subsection{Expansion of the soft mixed propagator}
\label{sec:outer}

Notice first that \eqref{seed_integral} is invariant under a simultaneous relabelling of the momentum ratios $r_{\alpha}$ and the masses $\mu_{\alpha}$, so without loss of generality we can take $r_1\leq r_2\leq r_3$. The mixed propagator we expand is then $\mathcal G_1$, the one with the smallest momentum, for reasons that will become clear shortly.

Writing both the Bessel functions and the ${}_2F_2$ term in \eqref{mixed_closed} as power series in $y$, and dropping the label $\alpha$ for the moment, the mixed propagator becomes
\begin{align}
\mathcal G(y)&=\sum_{n=0}^{\infty}\Big[\sum_{\pm}a^{\pm}_{n}\,y^{2n\pm i\mu}
 +b_{n}\,y^{n+1/2}\Big]\,, \label{mixed_expansion}
\end{align}
with the coefficients
\begin{align}
a^{\pm}_{n}&=\pm\frac{\pi^{3/2}\left(e^{\pm\pi\mu}-i\right)}{\sqrt2\,\sinh(2\pi\mu)}\,
 \frac{2^{-2n\mp i\mu}}{n!\,\Gamma(n+1\pm i\mu)}\,, \label{a_coeff}\\
b_{n}&=\frac{(-1)^{n+1}}{n!\,[(n+\frac12)^2+\mu^2]}
 \pFq[4]{3}{2}{1,-\frac n2,\frac{1-n}{2}}
 {\frac34-\frac{n+i\mu}2,\frac34-\frac{n-i\mu}2}{1}\,. \label{b_coeff}
\end{align}
The two terms capture the two types of small $y$ behaviour. The first one contains the nonanalytic powers $y^{2n\pm i\mu}$, while the second one contains only analytic powers of $y$, up to an overall factor of $y^{1/2}$. Note also that each $b_{n}$ is a finite sum, since the ${}_3F_2$ series terminates for every integer $n\geq0$.

We can now substitute \eqref{mixed_expansion} in \eqref{seed_integral}. Each term then becomes an integral of the product of two mixed propagators $\mathcal G_2\,\mathcal G_3$ weighted by a power of $Z$, and after rescaling $Z=y/r_3$ all of them are built from a single basic integral,
\begin{align}
Q_{23}(s,x)&=\int_0^\infty\dd y\,y^{s-1}\mathcal G_2(xy)\mathcal G_3(y)\,, \label{Q_def}
\end{align}
where $x\equiv r_2/r_3$. The integral converges for $\mathrm{Re}\,s>0$, a condition satisfied by all values of $s$ below. The bispectrum then reads
\begin{align}
\mathcal I&=r_3^{-3/2}\sum_{n=0}^{\infty}\bigg[b_{n}\Big(\frac{r_1}{r_3}\Big)^{n+1/2}Q_{23}(n+2,x)\nonumber\\
 &\quad+\sum_{\pm}a^{\pm}_{n}
 \Big(\frac{r_1}{r_3}\Big)^{2n\pm i\mu_1}Q_{23}\big(\tfrac32+2n\pm i\mu_1,x\big)\bigg]\,. \label{seed_expansion}
\end{align}
Let us now come back to the choice of $\mathcal G_1$. Although $\mathcal G_1$ itself decays at large $Z$, the nonanalytic and analytic series in \eqref{mixed_expansion} each grow as $e^{r_1Z}$, while $\mathcal G_2(r_2Z)\mathcal G_3(r_3Z)$ decays as $e^{-(r_2+r_3)Z}$. We can therefore exchange the sum and the integral as long as $r_1<r_2+r_3$, and our ordering guarantees this throughout the physical region. Notice also that $r_2\leq r_3$, together with the triangle inequality $r_3\leq r_1+r_2\leq2r_2$, restricts $x$ to $\frac12\leq x\leq1$. All that remains is to compute $Q_{23}(s,x)$ in this range, at $s=n+2$ and $s=\frac32+2n\pm i\mu_1$.

Actually, the integral \eqref{Q_def} is useful beyond the triple exchange considered here. For instance, in the double exchange diagram the external field attached to the cubic vertex simply replaces one of the mixed propagators by an exponential function, and expanding this exponential in powers of $Z$ once again leads to a series in $Q_{23}$. A detailed comparison with existing results for other tree diagrams will be presented in an upcoming work.

\subsection{Differential equation for the basic integral}
\label{sec:pair}

The basic integral \eqref{Q_def} is still not straightforward to compute directly. Instead, just as in the bootstrap approach to cosmological correlators, we can derive a differential equation for it. To see this, let us insert $\mathcal D_2$ acting on $\mathcal G_2$ into \eqref{Q_def}. This gives
\begin{align}
&\int_0^\infty\dd y\,y^{s-1}\mathcal G_3(y)\,(\mathcal D_2\mathcal G_2)(xy)\nonumber\\
&\qquad=\big[(x\partial_x)^2+\mu_2^2\big]Q_{23}(s,x)-x^2Q_{23}(s+2,x)\nonumber\\
&\qquad=-x^{1/2}\int_0^\infty\dd y\,y^{s-1/2}e^{-xy}\mathcal G_3(y)\,. \label{Q_eq2}
\end{align}
In the first equality we have used that $\mathcal G_2(xy)$ depends only on $xy$, so that the $y$ derivatives in $\mathcal D_2$ turn into $x$ derivatives acting on $Q_{23}$, while the $y^2$ term simply shifts $s$ by two. The second equality then follows from \eqref{mixed_eom}. In the same way, inserting $\mathcal D_3$ acting on $\mathcal G_3$ leads to
\begin{align}
&\int_0^\infty\dd y\,y^{s-1}\mathcal G_2(xy)\,(\mathcal D_3\mathcal G_3)(y)\nonumber\\
&\qquad=\big[(x\partial_x+s)^2+\mu_3^2\big]Q_{23}(s,x)-Q_{23}(s+2,x)\nonumber\\
&\qquad=-\int_0^\infty\dd y\,y^{s-1/2}e^{-y}\mathcal G_2(xy)\,. \label{Q_eq3}
\end{align}
This time the first equality requires an integration by parts, which moves the $y$ derivatives onto $y^{s-1}\mathcal G_2(xy)$ and turns them into $x\partial_x+s$, while the boundary terms vanish. Now notice that $Q_{23}(s+2,x)$ appears in both \eqref{Q_eq2} and \eqref{Q_eq3}. Eliminating it, we obtain a closed equation for $Q_{23}(s,x)$,
\begin{align}
&\big[\Delta_x+\mu_2^2-(s^2+\mu_3^2)x^2\big]Q_{23}(s,x)\nonumber\\
&=\int_0^\infty\dd y\,y^{s-1/2}\big[x^2e^{-y}\mathcal G_2(xy)-x^{1/2}e^{-xy}\mathcal G_3(y)\big]\,, \label{Q_ode}
\end{align}
with $\Delta_x\equiv x^2(1-x^2)\partial_x^2+x[1-(1+2s)x^2]\partial_x$. In practice, \eqref{Q_ode} is easier to solve after the change of variables
\begin{align}
 w&=1-x^2\,,
 \qquad
 Q_{23}(s,x)=x^{ i\mu_2}\mathcal Q(s,w)\,, \label{w_def}
\end{align}
under which the physical range $\frac12\leq x\leq1$ becomes $0\leq w\leq\frac34$. The equation then takes the simple form
\begin{align}
\big(\Delta_w-\gamma_+\gamma_-\big)\mathcal Q(s,w)&=\sum_{m=0}^{\infty}q_m(s)\,w^m\,, \label{Q_ode_w}
\end{align}
with $\Delta_w\equiv w(1-w)\partial_w^2+[s-(s+1+i\mu_2)w]\partial_w$ and $\gamma_\pm\equiv(s+i\mu_2\pm i\mu_3)/2$. Its homogeneous solutions are now just ${}_2F_1$ functions of $w$, and we have written the source as a power series in $w$, whose coefficients $q_m(s)$ will be determined next.

Let us now turn to the coefficients $q_m(s)$. After rescaling $y$ in the $\mathcal G_2$ term on the right hand side of \eqref{Q_ode}, $x$ enters the source integrals only through the exponentials. Since $w=0$ corresponds to $x=1$, we factor out their values at $x=1$, writing $e^{-xy}=e^{-y}e^{(1-x)y}$ and $e^{-y/x}=e^{-y}e^{(1-x^{-1})y}$, and then expand the second factors around $x=1$. In this way, all the $x$ dependence is moved outside the integrals, which take the form
\begin{align}
 \mathcal E_{\alpha}(a)&\equiv\int_0^\infty\dd y\,y^{a-1}e^{-y}\mathcal G_{\alpha}(y)\,, \label{E_def}
\end{align}
for different values of $a$. Since $\mathcal G_{\alpha}(y)$ stays bounded as $y\to0$, these integrals converge for $\mathrm{Re}\,a>0$, which is always the case here. The right hand side of \eqref{Q_ode_w} becomes
\begin{align}
\sum_{m=0}^{\infty}q_mw^m&=-\sum_{k=0}^{\infty}\frac{1}{4\,k!}
 \big[x^{-\frac32-i\mu_2}(1-x)^k\mathcal E_3(s+\tfrac12+k)\nonumber\\
&-x^{-\frac12-s-i\mu_2}(1-x^{-1})^k\mathcal E_2(s+\tfrac12+k)\big]\,. \label{source_series}
\end{align}
Because $1-x$ and $1-x^{-1}$ are both of order $w$, only $k\leq m$ contributes at order $w^m$, and expanding the remaining powers of $x=\sqrt{1-w}$ we finally obtain
\begin{align}
q_m(s)&=-\frac{1}{4\,m!}\sum_{k=0}^{m}\sum_{\ell=0}^{k}\frac{(-1)^\ell}{\ell!\,(k-\ell)!}\nonumber\\
&\quad\times\Big[\big(\tfrac34-\tfrac{\ell-i\mu_2}{2}\big)_m\mathcal E_3(s+\tfrac12+k) \label{q_closed}\\
&\quad-(-1)^k\big(\tfrac14+\tfrac{s+k-\ell+i\mu_2}{2}\big)_m\mathcal E_2(s+\tfrac12+k)\Big]\,.\nonumber
\end{align}
Here $(\cdot)_m$ is the Pochhammer symbol. It remains to compute \eqref{E_def}, which turns out to have a simple physical interpretation. Since each mixed propagator already contains one mixing vertex, \eqref{E_def} is nothing else but the two-point function from the single massive exchange, in which the other vertex carries a general power of the conformal time. We can then directly borrow the known results for such two-point functions with arbitrary vertex powers \cite{Qin:2023ejc}. Translating them into our notation yields
\begin{align}
\mathcal E_{\alpha}(a)&=-\frac{\sqrt\pi\,\beta_{\alpha}\,\Gamma(a+i\mu_{\alpha})\Gamma(a-i\mu_{\alpha})}{2^a\,\Gamma(a+\frac12)}\nonumber\\
 &+\frac{\Gamma(a+\frac12)}{2^{a+1/2}(\frac12+i\mu_{\alpha})^2}
 \pFq{3}{2}{1,1,1+i\mu_{\alpha}-a}{\frac32+i\mu_{\alpha},\frac32+i\mu_{\alpha}}{1}\,.\label{E_closed}
\end{align}

Because the basic integral \eqref{Q_def} is analytic around $x=1$, the solution of \eqref{Q_ode_w} can be expressed as the power series in $w$, $\mathcal Q(s,w)=\sum_m c_m(s)w^m$. Plugging it into \eqref{Q_ode_w}, we find that the coefficients satisfy the recurrence relation
\begin{align}
c_{m+1}(s)&=\frac{(m+\gamma_+)(m+\gamma_-)\,c_m(s)+q_m(s)}{(m+1)(m+s)}\,. \label{c_recurrence}
\end{align}
This recurrence relation can be solved directly in terms of the initial value $c_0$. For $m\geq1$, we find
\begin{align}
c_m&=\frac{(\gamma_+)_m(\gamma_-)_m}{(s)_m\,m!}\bigg[c_0+\sum_{n=0}^{m-1}\frac{(s)_n\,n!\,q_n}{(\gamma_+)_{n+1}(\gamma_-)_{n+1}}\bigg]\,. \label{c_closed}
\end{align}
Summing over $m$ then gives
\begin{align}
\mathcal Q&=c_0\,\pFq{2}{1}{\gamma_+,\gamma_-}{s}{w}+\sum_{n=0}^{\infty}\frac{q_n\,w^{n+1}}{(n+1)(n+s)}\nonumber\\
&\quad\times\pFq{3}{2}{1,\gamma_++n+1,\gamma_-+n+1}{n+2,s+n+1}{w}\,. \label{Q_closed}
\end{align}
The first term is the homogeneous solution, and the second one comes from the source. Since $q_n$ is already known from \eqref{q_closed}, the only unknown part is $c_0$, which we compute in the next subsection.

\subsection{Boundary value from equal momentum}
\label{sec:equal}

Setting $w=0$ in \eqref{Q_closed}, we see that the boundary value $c_0$ is simply the basic integral at equal momentum $k_2=k_3$,
\begin{align}
c_0(s)&=Q_{23}(s,1)=\int_0^\infty\dd y\,y^{s-1}\mathcal G_2(y)\mathcal G_3(y)\,. \label{c0_def}
\end{align}
The integrand is still a product of two mixed propagators, but now with the same argument. Such a product is most easily handled with the Laplace representation recently developed in \cite{Belrhali:2026ygh,Belrhali:2026jqe,Belrhali:2026uxn}, in which each mixed propagator becomes a superposition of exponentials,
\begin{align}
\mathcal G_{\alpha}(y)&=\sqrt y\int_1^\infty\dd t\,e^{-yt}\widetilde{\mathcal G}_{\alpha}(t)\,,\nonumber\\
\widetilde{\mathcal G}_{\alpha}(t)&=Q_{-\frac12+i\mu_{\alpha}}(t)-\sqrt{\frac\pi2}\,\beta_{\alpha}P_{-\frac12+i\mu_{\alpha}}(t)\,, \label{laplace_rep}
\end{align}
with $P_\nu$ and $Q_\nu$ the Legendre functions of the first and second kind. The $y$ integral in \eqref{c0_def} then becomes elementary, and we obtain
\begin{align}
c_0(s)&=\Gamma(s+1)\int_1^\infty\dd t_2\int_1^\infty\dd t_3\,\frac{\widetilde{\mathcal G}_2(t_2)\widetilde{\mathcal G}_3(t_3)}{(t_2+t_3)^{s+1}}\,. \label{c0_laplace}
\end{align}
The two $t$ integrals are coupled only through the denominator $(t_2+t_3)^{s+1}$. To separate them, we use
\begin{align}
t_2+t_3&=\frac{(t_2+1)(t_3+1)}{2}
\left[1-\frac{(t_2-1)(t_3-1)}{(t_2+1)(t_3+1)}\right]\,. \label{factorisation}
\end{align}
The ratio in the square brackets lies in $[0,1)$, so we can expand $(t_2+t_3)^{-s-1}$ in powers of it. Each term factorises into two one-dimensional integrals, giving
\begin{align}
c_0(s)&=\frac{\Gamma(s+1)}{2^{s+1}}\sum_{n=0}^{\infty}\frac{(s+1)_n}{n!}\prod_{\alpha=2}^{3}\mathcal W_{\alpha}(s,n)\,, \label{c0_series}
\end{align}
with
\begin{align}
\mathcal W_{\alpha}(s,n)&=2^{s+1}\int_1^\infty\dd t\,\widetilde{\mathcal G}_{\alpha}(t)\,\frac{(t-1)^n}{(t+1)^{n+s+1}}\,. \label{W_def}
\end{align}
The integrals in \eqref{W_def} can be carried out once the Legendre functions are written in terms of ${}_2F_1$, and the result reads
\begin{align}
&\mathcal W_{\alpha}=\frac{(n!)^2}{(\frac12+i\mu_{\alpha})_{n+1}^2}\,\pFq[4]{3}{2}{n+1,n+1,\frac12+i\mu_{\alpha}-s}{n+\frac32+i\mu_{\alpha},n+\frac32+i\mu_{\alpha}}{1}\nonumber\\
&\enspace-\frac{\sqrt{2\pi}\,\beta_{\alpha}\,n!}{(s+\frac12-i\mu_{\alpha})_{n+1}}\,\pFq[4]{3}{2}{\frac12-i\mu_{\alpha},\frac12-i\mu_{\alpha},n+1}{1,n+s+\frac32-i\mu_{\alpha}}{1}\,. \label{W_closed}
\end{align}
Both ${}_3F_2$ functions converge for $\mathrm{Re}\,s>-\frac12$, while the sum in \eqref{c0_series} converges for $\mathrm{Re}\,s>0$. These conditions are satisfied by all the values of $s$ appearing in \eqref{seed_expansion}.
\\
\\
The derivation above has involved several steps. Let us pause here and collect the final result.
\begin{tcolorbox}[colback=white,colframe=black!45,boxrule=0.4pt,sharp corners,left=5pt,right=5pt,top=4pt,bottom=4pt]
\noindent\textbf{Summary of the result.} Given the masses $\mu_\alpha$ and the momentum ratios $r_1\leq r_2\leq r_3$, we write $x=r_2/r_3$, $w=1-x^2$ and $\gamma_\pm=(s+i\mu_2\pm i\mu_3)/2$. The seed integral then takes the form of a double series,
\begin{align}
&\mathcal I=x^{i\mu_2}\sum_{n=0}^{\infty}\sum_{s}\mathcal A_s\,r_1^{s-\frac32}r_3^{-s}\bigg\{c_0(s)\,\pFq{2}{1}{\gamma_+,\gamma_-}{s}{w}\nonumber\\
&\quad+\sum_{m=0}^{\infty}\frac{q_m(s)\,w^{m+1}}{(m+1)(m+s)}\nonumber\\
&\quad\times\pFq{3}{2}{1,\gamma_++m+1,\gamma_-+m+1}{m+2,s+m+1}{w}\bigg\}\,. \label{full_result}
\end{align}
\end{tcolorbox}
\noindent Here $s$ takes the three values $n+2$ and $\frac32+2n\pm i\mu_1$, with $\mathcal A_{n+2}=b_n$ and $\mathcal A_{\frac32+2n\pm i\mu_1}=a_n^\pm$. The ingredients are
\begin{enumerate}[label=(\roman*), leftmargin=2.5em]
\item the coefficients $a_n^\pm$ and $b_n$ in \eqref{a_coeff} and \eqref{b_coeff},
\item the source coefficients $q_m$ in \eqref{q_closed}, with $\mathcal E_\alpha$ in \eqref{E_closed},
\item the boundary value $c_0$ in \eqref{c0_series}, with $\mathcal W_\alpha$ in \eqref{W_closed}.
\end{enumerate}
These ingredients are expressed in terms of Gamma and generalised hypergeometric functions, so the result can be evaluated directly. For isosceles configurations with $k_2=k_3$, where $w=0$, the sum over $m$ in \eqref{full_result} drops out, and the result reduces to a single sum that requires only the boundary values $c_0(s)$. 

In practice, evaluating the ${}_3F_2$ functions at unity and the series \eqref{c0_series} term by term can be time consuming, and it is more efficient to generate the required quantities directly from recurrence relations such as \eqref{c0_recurrence} and \eqref{c_recurrence}.

\begin{figure*}[t]
\centering
\includegraphics[width=\textwidth]{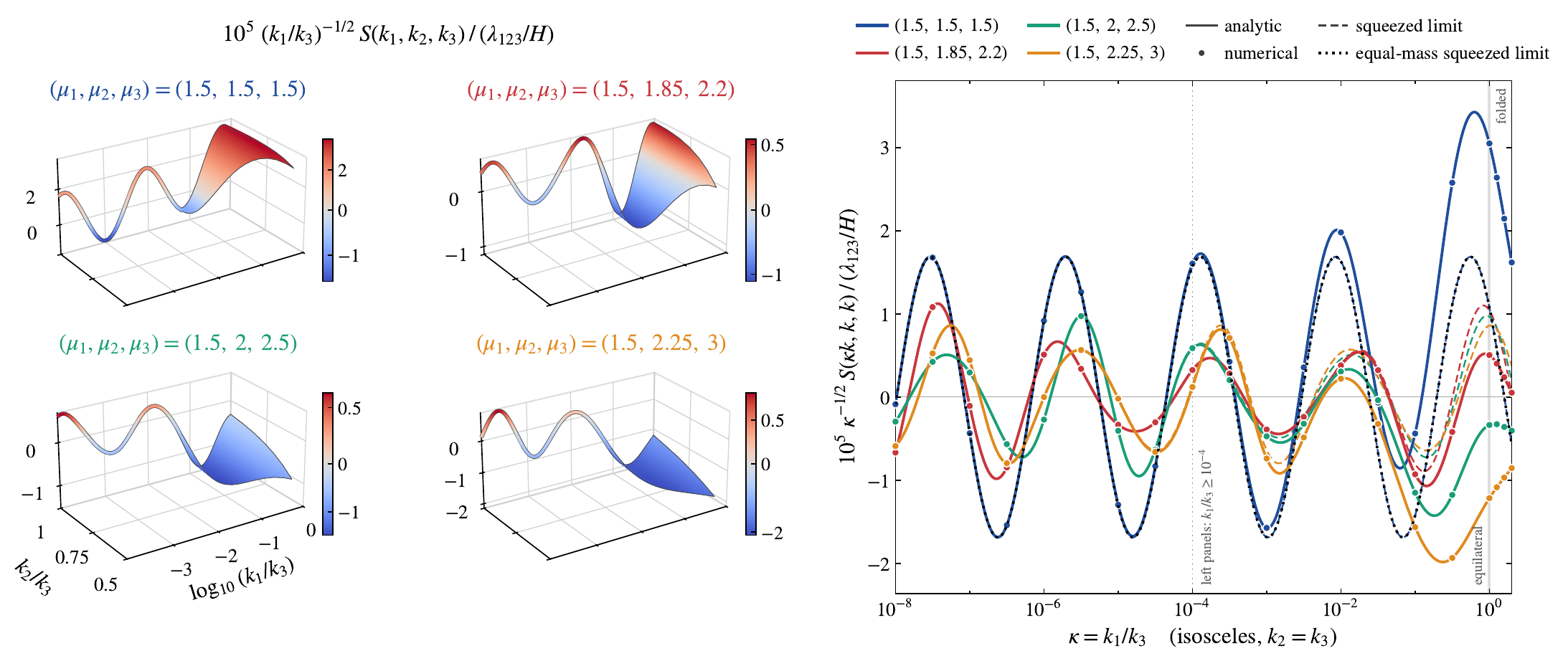}
\caption{Triple exchange bispectrum for three different massive fields whose only nonzero cubic couplings are $\lambda_{123}$ and its permutations, so that the sum~\eqref{full_bispectrum} runs over the six ways of attaching the three fields to the external legs. We define $S(k_1,k_2,k_3)\equiv(k_1k_2k_3)^2\langle\varphi_{\boldsymbol k_1}\varphi_{\boldsymbol k_2}\varphi_{\boldsymbol k_3}\rangle'/H^3$ and multiply it by $(k_1/k_3)^{-1/2}$ to remove its overall decay towards the squeezed limit. The result is shown in units of $10^{-5}\lambda_{123}/H$ for four sets of mass parameters $(\mu_1,\mu_2,\mu_3)$ that share the same lightest field. The mixings $\rho_\alpha=(\tfrac14+\mu_\alpha^2)H/20$ offset the suppression by $1/(\tfrac14+\mu_\alpha^2)$ of each massive line and give the four sets a similar size. Since the couplings only enter through the overall factor $\lambda_{123}\rho_1\rho_2\rho_3$, any other choice rescales each set. \emph{Left:} shape over all triangles with $k_1\leq k_2\leq k_3$ and $k_1/k_3\geq10^{-4}$, as a function of $\log_{10}(k_1/k_3)$ and $k_2/k_3$, from the analytical result~\eqref{seed_expansion} on about $10^3$ triangles per panel. Each panel has its own vertical and colour scale, with white at zero. The back edge of each surface is the isosceles line $k_2=k_3$, which ends at the equilateral configuration, while the front edge consists of the folded configurations $k_1+k_2=k_3$ for $k_1/k_3\leq\tfrac12$ and of the isosceles configurations $k_1=k_2$ beyond. \emph{Right:} the same four sets, in the same colours, along the isosceles line $k_2=k_3$ as a function of $\kappa=k_1/k_3$, from $\kappa=10^{-8}$ to the folded configuration $\kappa=2$. The dotted vertical line marks the lower end of the left panels, and the grey band the equilateral configuration. Solid lines: analytical result~\eqref{seed_expansion}. Dots: numerical integration of~\eqref{seed_integral}, which agrees with the analytical result to numerical precision. Dashed lines: leading squeezed limit~\eqref{squeezed}-\eqref{Z_def}, drawn over the whole range to show where it sets in. Dotted black line: the same limit for equal masses from the closed form~\eqref{Z_equal}-\eqref{c0tilde}. For equal masses, the signal oscillates in $\log\kappa$ at a single frequency with a constant amplitude. For different masses, each field contributes an oscillation at a frequency set by its own mass, and their superposition is a modulated oscillation: the closer the masses, the deeper and the slower the modulation.}
\label{fig:shapes}
\end{figure*}

Finally, we show in the left panel of Fig.~\ref{fig:shapes} the full shapes for four sets of masses that share the same lightest field, with the quadratic mixings $\rho_\alpha=(\frac14+\mu_\alpha^2)H/20$ chosen to offset the suppression of heavier fields. Since \eqref{full_result} converges throughout the physical region, the same expression is used for every triangle, from squeezed to folded configurations.

\section{Equal masses and squeezed limit}
\label{sec:kinematics}

Our result also gives direct access to the squeezed limit $\kappa\equiv k_1/k_3\to0$. In this limit $x\to1$, so that each basic integral in \eqref{seed_expansion} reduces to its boundary value $c_0$, while only the $n=0$ terms contribute at leading order. This gives
\begin{align}
\mathrm{Re}\,\mathcal I&=|\mathcal Z|\cos\big(\mu_1\log\kappa+\arg\mathcal Z\big)+\mathcal O(\kappa^{1/2})\,, \label{squeezed}
\end{align}
and here
\begin{align}
\mathcal Z&\equiv a_0^+c_0\big(\tfrac32+i\mu_1\big)+\big[a_0^-c_0\big(\tfrac32-i\mu_1\big)\big]^*\,. \label{Z_def}
\end{align}
We therefore see that the frequency of the oscillation in $\log\kappa$ is fixed by the mass of the field carrying the soft momentum, while the other two masses only affect its amplitude and phase through $c_0(\frac32\pm i\mu_1)$. Once the sum in \eqref{full_bispectrum} is carried out, each exchanged field gives rise to its own oscillation, with a frequency set by its mass. Note that this structure cannot be seen for a single massive species, for which all oscillations share the same frequency.

When the masses coincide, the amplitude $\mathcal Z$ can be computed in closed form, which also allows for a nontrivial comparison with the existing result. Let us therefore set $\mu_1=\mu_2=\mu_3\equiv\mu$ and drop the index $\alpha$. The two factors $\mathcal W_\alpha$ in \eqref{c0_series} then coincide, and the boundary value reads
\begin{align}
c_0(s)&=\frac{\Gamma(s+1)}{2^{s+1}}\sum_{n=0}^{\infty}\frac{(s+1)_n}{n!}\,\mathcal W(s,n)^2\,. \label{c0_equal}
\end{align}
At $s=\frac32+i\mu$, which is the value entering \eqref{Z_def}, the first ${}_3F_2$ in \eqref{W_closed} terminates after two terms, and we find
\begin{align}
&\mathcal W\big(\tfrac32+i\mu,n\big)=\frac{(n!)^2(\frac12+i\mu)(2n+\frac52+i\mu)}{(\frac12+i\mu)_{n+2}^2}\nonumber\\
&\quad-\frac{\sqrt{2\pi}\,\beta}{(n+1)_2}\,\pFq[4]{3}{2}{\frac12-i\mu,\frac12-i\mu,n+1}{1,n+3}{1}\,. \label{W_squeezed}
\end{align}

The other value, $c_0(\frac32-i\mu)$, does not truncate, but it is related to $c_0(\frac32+i\mu)$. On the real axis, $I_{i\mu}^*=I_{-i\mu}$ and the ${}_2F_2$ term in \eqref{mixed_closed} is real, so that complex conjugation only changes the coefficient of $K_{i\mu}$, from $\beta$ to $\tilde\beta\equiv\beta-i\sqrt{2\pi}$. Because this coefficient appears only in the second term of \eqref{W_closed}, we find
\begin{align}
c_0\big(\tfrac32-i\mu\big)^*&=c_0\big(\tfrac32+i\mu\big)\Big|_{\beta\to\tilde\beta}\,. \label{c0_conj}
\end{align}
On the other hand, \eqref{c0_equal} and \eqref{W_squeezed} show that $c_0(\frac32+i\mu)$ is a quadratic polynomial in $\beta$, which we write as $c_0(\frac32+i\mu)=\tilde c_0+\beta\tilde c_1+\beta^2\tilde c_2$, with $\tilde c_k$ the coefficient of $\beta^k$. Inserting both into \eqref{Z_def}, the term proportional to $\tilde c_1$ comes with $a_0^+\beta+a_0^{-*}\tilde\beta$, which vanishes for the coefficients \eqref{a_coeff}, and we obtain
\begin{align}
\mathcal Z&=(a_0^++a_0^{-*})\,\tilde c_0+(a_0^+\beta^2+a_0^{-*}\tilde\beta^2)\,\tilde c_2 \label{Z_equal}\\
&=\frac{\sqrt2\,\pi^{3/2}}{2^{i\mu}\Gamma(1+i\mu)\sinh(2\pi\mu)}\big[(\sinh\pi\mu-i)\tilde c_0+i\pi\tilde c_2\big]\,. \nonumber
\end{align}

Both $\tilde c_0$ and $\tilde c_2$ can in fact be evaluated in closed form. Let us start with $\tilde c_2$. Since $\beta^2$ multiplies only the second term of \eqref{W_squeezed}, which originates from the $K_{i\mu}$ term in \eqref{mixed_closed}, $\tilde c_2$ is nothing but the integral \eqref{c0_def} with both mixed propagators replaced by $K_{i\mu}$. This integral is known in closed form \cite{Gradshteyn2007} which reads
\begin{align}
\tilde c_2&=\int_0^\infty\dd y\,y^{\frac12+i\mu}K_{i\mu}(y)^2\nonumber\\
&=\frac{\sqrt\pi\,\Gamma(\frac34+\frac{i\mu}2)\Gamma(\frac34-\frac{i\mu}2)\Gamma(\frac34+\frac{3i\mu}2)}{4\,\Gamma(\frac54+\frac{i\mu}2)}\,. \label{c2tilde}
\end{align}
For $\tilde c_0$, only the first term of \eqref{W_squeezed} contributes. Inserting it into \eqref{c0_equal} at $s=\frac32+i\mu$, we find
\begin{align}
\tilde c_0&=\frac{\Gamma(\frac12+i\mu)}{2^{\frac12+i\mu}(\frac12+i\mu)(\frac32+i\mu)^3}\sum_{n=0}^{\infty}\frac{(n!)^3\,(n+\frac54+\frac{i\mu}2)^2}{(\frac52+i\mu)_n^3}\,. \label{c0tilde_sum}
\end{align}
This sum can be carried out, giving
\begin{align}
\tilde c_0&=\frac{\Gamma(\frac12+i\mu)}{2^{\frac12+i\mu}(\frac12+i\mu)^2}\bigg[\frac13-\frac1{12}\Big(\frac{\frac12+i\mu}{\frac32+i\mu}\Big)^3\zeta_3\big(\tfrac52+i\mu\big)\bigg]\,, \label{c0tilde}
\end{align}
with $\zeta_3(a)\equiv{}_4F_3(1,1,1,1;a,a,a;1)$, which reduces to Ap\'ery's constant $\zeta(3)$ at $a=2$. Together with \eqref{Z_equal} and \eqref{c2tilde}, this gives the squeezed limit for equal masses in closed form. It agrees with the result of \cite{Pinol:2026hey}, which was obtained with a completely different technique.

In the right panel of Fig.~\ref{fig:shapes}, we compare the behaviour of the general analytical formula in the squeezed limit with the one obtained from \eqref{squeezed}-\eqref{Z_def}, and from \eqref{Z_equal}-\eqref{c0tilde} in the equal mass case, for the same masses and quadratic mixings as in the left panel.

\section{Summary}
\label{sec:summary}

We have derived an analytical result for the triple massive exchange bispectrum generated by a cubic interaction among three scalar fields with generally different masses, working at leading order in the quadratic mixings. The final result is a double series involving generalised hypergeometric functions of a single variable, which converges throughout the whole physical region, including squeezed and folded configurations.

The key step is to expand the mixed propagator carrying the smallest external momentum. This reduces the triple exchange integral to a family of integrals involving the other two mixed propagators. The equation satisfied by the mixed propagators then yields an inhomogeneous differential equation in a momentum ratio. Analyticity at equal momentum selects the appropriate solution, while a Laplace representation determines the remaining boundary value analytically. All coefficients in the resulting series are therefore specified explicitly, with no unevaluated time integrals or undetermined boundary data.

Our result also gives direct access to the squeezed limit. For equal masses, the leading oscillatory coefficient simplifies to a very compact form without any summations, reproducing the result obtained with a completely different method. We have also checked the analytical results against numerical calculations and found perfect agreement for the mass parameters and momentum configurations tested.

An interesting direction is to extend the present framework to nonperturbative quadratic mixing involving several massive fields, with the aim of obtaining an explicit analytical series that converges throughout the physical region and involves no unevaluated integrals. The results also motivate a detailed phenomenological study of triple exchange with different masses, including its distinguishability from single-species triple exchange and from single- and double-exchange contributions, as well as the construction of templates for observational searches.

\vspace{3pt}
\paragraph*{Note added} While we were preparing this manuscript, two works on the triple exchange of a single massive species appeared. In \cite{Chen:2026khu}, the bispectrum was expressed as a convergent double series at weak mixing, using the Mellin-Barnes representation, and the approach was extended to general mixing strength. The same diagram was studied in \cite{Ahmed:2026qtc} for light fields at weak mixing, by expanding the bispectrum in the mass around the massless and conformally coupled limits and expressing the momentum dependence of each coefficient in terms of multiple polylogarithms.

\vspace{3pt}
\paragraph*{Acknowledgements}
We thank Denis Werth and Zhong-Zhi Xianyu for useful discussions. S.A. is supported by the Japan Science and Technology Agency (JST) as part of Adopting Sustainable Partnerships for Innovative Research Ecosystem (ASPIRE), grant JPMJAP2318, and by JSPS KAKENHI Grant Number 26K07078. Z.Q. is supported by the China Scholarship Council (CSC). F.S. is supported by JSPS KAKENHI Grant Number 26KJ0060. M.Y. is supported by IBS under the project code, IBS-R018-D3, and by JSPS Grant-in-Aid for Scientific Research Number JP23K20843. Y.Z. is supported by IBS under the project code, IBS-R018-D3. The authors used ChatGPT (OpenAI, GPT-5.6 Sol and GPT-6 Astra) and Claude (Anthropic, Fable 5.1) to assist in checking derivations, identifying possible inconsistencies, and improving the presentation of the manuscript. All AI-assisted outputs were independently checked by the authors, who take full responsibility for the content.

\bibliographystyle{utphys}
\bibliography{references}

\clearpage
\onecolumngrid
\appendix
\setcounter{secnumdepth}{2}

\section{Continuation to arbitrary positive masses}
\label{app:complementary}

The results above extend to arbitrary positive masses. For an internal line with $0<m_\alpha<3H/2$ we choose $\mu_\alpha=-i\sqrt{9/4-m_\alpha^2/H^2}$, so that $-3/2<\mathrm{Im}\,\mu_\alpha\leq0$ for every positive mass. With this choice, the result \eqref{full_result} and the ingredients (i) and (ii) listed below it apply directly. However, for the ingredient (iii) the series \eqref{c0_series} for $c_0(s)$ may diverge at some of the values of $s$ entering \eqref{seed_expansion}.
This does not by itself imply a divergence of the physical bispectrum. For the cubic interaction considered here, spurious late-time divergences in individual contributions cancel in the full in-in correlator, which remains finite for strictly positive internal masses \cite{Chen:2009zp}.

To determine the convergence of \eqref{c0_series}, we need the large $n$ behaviour of $\mathcal W_\alpha(s,n)$ in \eqref{W_closed}. Applying Thomae's transformation to its first term, and the Euler integral representation followed by the Gauss connection formula to its second, gives \cite{NIST:DLMF}
\begin{align}
 \lim_{n\to\infty}\mathcal W_\alpha(s,n)\sim \sum_{\pm}a_0^\pm\,
 \frac{2^{1/2\mp i\mu_\alpha}\Gamma(s+\tfrac12\pm i\mu_\alpha)}
 {\Gamma(\tfrac12\mp i\mu_\alpha)}\,
 n^{-s-1/2\mp i\mu_\alpha}\,,
\end{align}
with $a_0^\pm$ given by \eqref{a_coeff} at $\mu=\mu_\alpha$, at fixed $s$. The modulus $|\mathcal W_\alpha(s,n)|$ thus falls off as $n^{-\mathrm{Re}\,s-1/2-\mathrm{Im}\,\mu_\alpha}$. Together with $(s+1)_n/n!\sim n^s/\Gamma(s+1)$, the $n$-th term of \eqref{c0_series} then scales as $n^{-1-\mathrm{Re}\,s-\mathrm{Im}(\mu_2+\mu_3)}$, so that the series converges absolutely for
\begin{align}
 \mathrm{Re}\,s>-\mathrm{Im}(\mu_2+\mu_3)\,.
 \label{c0_convergence}
\end{align}
This is also the condition for the integral \eqref{c0_def} to converge at $y=0$. However, the smallest value of $\mathrm{Re}\,s$ in \eqref{seed_expansion}, namely $\tfrac32+\mathrm{Im}\,\mu_1$, need not satisfy \eqref{c0_convergence}.

This can be cured by a recurrence relating $c_0(s)$ to $c_0(s+2)$ and the integrals $\mathcal E_\alpha$. For positive masses, every $s$ in \eqref{seed_expansion} has $\mathrm{Re}\,s>0$, while $-\mathrm{Im}(\mu_2+\mu_3)<3$, so that $s+4$ always satisfies \eqref{c0_convergence}. We can therefore evaluate $c_0(s+4)$ from the convergent series \eqref{c0_series} and apply the recurrence twice to recover $c_0(s)$.

To derive the recurrence, we start from the identity
\begin{align}
 2\int_0^\infty\dd y\,y^{s-1}
 (y\partial_y\mathcal G_2)(y\partial_y\mathcal G_3)
 ={}&\int_0^\infty\dd y\,y^{s-1}
 (y\partial_y)^2(\mathcal G_2\mathcal G_3)-\int_0^\infty\dd y\,y^{s-1}
 \big[\mathcal G_3(y\partial_y)^2\mathcal G_2
 +\mathcal G_2(y\partial_y)^2\mathcal G_3\big]\,.
\label{product_ibp}
\end{align}
Using integration by parts (IBP) twice, with $\mathrm{Re}\,s$ large enough for the boundary terms to vanish, the first integral on the right hand side becomes
\begin{align}
 \int_0^\infty\dd y\,y^{s-1}(y\partial_y)^2(\mathcal G_2\mathcal G_3)
 =-s\int_0^\infty\dd y\,y^{s-1}(y\partial_y)(\mathcal G_2\mathcal G_3)
 =s^2c_0(s)\,,
\end{align}
as follows from \eqref{c0_def}. For the second integral, together with its minus sign, the differential equation \eqref{mixed_eom} gives
\begin{align}
 &-\int_0^\infty\dd y\,y^{s-1}
 \big[\mathcal G_3(y\partial_y)^2\mathcal G_2
 +\mathcal G_2(y\partial_y)^2\mathcal G_3\big]\nonumber\\
 &\quad=-\int_0^\infty\dd y\,y^{s-1}
 \big[(2y^2-\mu_2^2-\mu_3^2)\mathcal G_2\mathcal G_3
 -\sqrt y\,e^{-y}(\mathcal G_2+\mathcal G_3)\big]\nonumber\\
 &\quad=(\mu_2^2+\mu_3^2)\int_0^\infty\dd y\,y^{s-1}\mathcal G_2\mathcal G_3
 -2\int_0^\infty\dd y\,y^{s+1}\mathcal G_2\mathcal G_3
 +\sum_{\alpha=2}^3\int_0^\infty\dd y\,y^{s-1/2}e^{-y}\mathcal G_\alpha\nonumber\\
 &\quad=(\mu_2^2+\mu_3^2)c_0(s)-2c_0(s+2)
 +\sum_{\alpha=2}^3\mathcal E_\alpha(s+\tfrac12)\,,
\end{align}
where in the final equality we have used \eqref{c0_def} and \eqref{E_def}. Multiplying the left hand side of \eqref{product_ibp} by $-s$ and using IBP together with \eqref{mixed_eom}, we obtain
\begin{align}
 &-2s\int_0^\infty\dd y\,y^{s-1}
 (y\partial_y\mathcal G_2)(y\partial_y\mathcal G_3)
 =2\int_0^\infty\dd y\,y^{s-1}(y\partial_y)
 \big[(y\partial_y\mathcal G_2)(y\partial_y\mathcal G_3)\big]\nonumber\\
 ={}&\int_0^\infty\dd y\,y^{s-1}
 \Big[(2y^2-\mu_2^2-\mu_3^2)(y\partial_y)(\mathcal G_2\mathcal G_3)+(\mu_2^2-\mu_3^2)
 \big[(y\partial_y\mathcal G_2)\mathcal G_3
 -\mathcal G_2(y\partial_y\mathcal G_3)\big]-2\sqrt y\,e^{-y}(y\partial_y)(\mathcal G_2+\mathcal G_3)\Big]\nonumber\\
 ={}&-2(s+2)c_0(s+2)+s(\mu_2^2+\mu_3^2)c_0(s)
 +2\sum_{\alpha=2}^3
 \big[(s+\tfrac12)\mathcal E_\alpha(s+\tfrac12)-\mathcal E_\alpha(s+\tfrac32)\big]\nonumber\\
 &+(\mu_2^2-\mu_3^2)\int_0^\infty\dd y\,y^{s-1}
 \big[(y\partial_y\mathcal G_2)\mathcal G_3
 -\mathcal G_2(y\partial_y\mathcal G_3)\big]\,.
 \label{derivative_ibp}
\end{align}
Here the terms proportional to $\sqrt y\,e^{-y}$ also follow from IBP,
\begin{align}
 \int_0^\infty\dd y\,y^{s-1}\sqrt y\,e^{-y}
 (y\partial_y\mathcal G_\alpha)
 =-(s+\tfrac12)\mathcal E_\alpha(s+\tfrac12)+\mathcal E_\alpha(s+\tfrac32)\,.
\end{align}
The remaining integral in \eqref{derivative_ibp} is obtained in the same way. Acting with $y\partial_y$ on its antisymmetric integrand, the $y^2$ terms cancel by \eqref{mixed_eom}, and IBP then gives
\begin{align}
 s\int_0^\infty\dd y\,y^{s-1}
 \big[(y\partial_y\mathcal G_2)\mathcal G_3-\mathcal G_2(y\partial_y\mathcal G_3)\big]
 =(\mu_2^2-\mu_3^2)c_0(s)-4q_0(s)\,,
\end{align}
with $4q_0(s)=\mathcal E_2(s+\tfrac12)-\mathcal E_3(s+\tfrac12)$ from \eqref{q_closed}. Combining these relations to eliminate both derivative integrals, and multiplying by $s$, we arrive at
\begin{align}
 \big[s^2+(\mu_2+\mu_3)^2\big]\big[s^2+(\mu_2-\mu_3)^2\big]c_0(s)
 ={}&4s(s+1)c_0(s+2)+4(\mu_2^2-\mu_3^2)q_0(s)\nonumber\\
 &+s\sum_{\alpha=2}^3
 \big[2\mathcal E_\alpha(s+\tfrac32)-(3s+1)\mathcal E_\alpha(s+\tfrac12)\big]\,,
\label{c0_recurrence}
\end{align}
which is symmetric under $2\leftrightarrow3$ as expected, since both $q_0(s)$ and $\mu_2^2-\mu_3^2$ change sign.

Solving \eqref{c0_recurrence} for $c_0(s)$ then continues it to the larger domain where the series for $c_0(s+2)$ converges, with $\mathcal E_\alpha$ given by \eqref{E_closed}. Writing
\begin{align}
 R(s)&=\frac{4s(s+1)}
 {\big[s^2+(\mu_2+\mu_3)^2\big]\big[s^2+(\mu_2-\mu_3)^2\big]}\,,\\
 S(s)&=\frac{(\mu_2^2-\mu_3^2)[\mathcal E_2(s+\tfrac12)-\mathcal E_3(s+\tfrac12)]
 +s\displaystyle\sum_{\alpha=2}^3\big[2\mathcal E_\alpha(s+\tfrac32)-(3s+1)\mathcal E_\alpha(s+\tfrac12)\big]}
 {\big[s^2+(\mu_2+\mu_3)^2\big]\big[s^2+(\mu_2-\mu_3)^2\big]}\,,
\end{align}
we then have the recurrence relation
\begin{align}
c_0(s)&=R(s)c_0(s+2)+S(s)\,.
\end{align}
Applying this twice gives
\begin{align}
c_0(s)=R(s)R(s+2)c_0(s+4)+R(s)S(s+2)+S(s)\,,
\end{align}
with $c_0(s+4)$ given by \eqref{c0_series}. The resulting $c_0(s)$ is then inserted into \eqref{Q_closed} and \eqref{seed_expansion}, with all other dependence on $s$ unchanged.

\end{document}